\documentclass[twocolumn]{aa}
\usepackage[varg]{txfonts}
\usepackage{courier}
\usepackage[T1]{fontenc}
\usepackage{gensymb}
\usepackage{amssymb}
\usepackage{amsmath}
\usepackage{url}
\DeclareUnicodeCharacter{2212}{-}

\usepackage{mathtools}
\usepackage{changepage} 
\usepackage{url}
\usepackage{appendix}
\usepackage{xcolor}
\usepackage{xspace}

\newcommand{\cha}{\textit{Chandra}\xspace}
\newcommand{\bat}{\textit{Swift}-BAT\xspace}

\newcommand{\xmm}{\textit{XMM-Newton}\xspace}
\newcommand{\xrt}{\textit{Swift}-XRT\xspace}

\begin{document}

\title{A New Mid-Infrared and X-ray Machine Learning Algorithm to Discover Compton-thick AGN}

\author{R. Silver\inst{1} \and N. Torres-Alb\`{a}\inst{1} \and X. Zhao\inst{2, 1} \and S. Marchesi\inst{3, 1} \and A. Pizzetti\inst{1} \and I.Cox\inst{1} \and M. Ajello\inst{1}}

\institute{Department of Physics and Astronomy, Clemson University,  Kinard Lab of Physics, Clemson, SC 29634, USA
\and
Harvard-Smithsonian Center for Astrophysics, 60 Garden Street, Cambridge, MA 02138, USA
\and
INAF - Osservatorio di Astrofisica e Scienza dello Spazio di Bologna, Via Piero Gobetti, 93/3, 40129, Bologna, Italy}

\abstract
{We present a new method to predict the line-of-sight column density (N$_{\rm H}$) values of active galactic nuclei (AGN) based on mid-infrared (MIR), soft, and hard X-ray data. We developed a multiple linear regression machine learning algorithm trained with WISE colors, \textit{Swift}-BAT count rates, soft X-ray hardness ratios, and an MIR$-$soft X-ray flux ratio. Our algorithm was trained off 451 AGN from the \textit{Swift}-BAT sample with known N$_{\rm H}$ and has the ability to accurately predict N$_{\rm H}$ values for AGN of all levels of obscuration, as evidenced by its Spearman correlation coefficient value of 0.86 and its 75\% classification accuracy. This is significant as few other methods can be reliably applied to AGN with Log(N$_{\rm H} <$) 22.5. It was determined that the two soft X-ray hardness ratios and the MIR$-$soft X-ray flux ratio were the largest contributors towards accurate N$_{\rm H}$ determination. This algorithm will contribute significantly to finding Compton-thick (CT-) AGN (N$_{\rm H} \geq$ 10$^{24}$ cm$^{-2}$), thus enabling us to determine the true intrinsic fraction of CT-AGN in the local universe and their contribution to the Cosmic X-ray Background.
}

\keywords{Galaxies: active -- Galaxies: nucleus -- Infrared: galaxies
-- X-rays: galaxies}

\maketitle

\section{Introduction}
Active Galactic Nuclei (AGN) are supermassive black holes (SMBHs) that reside in the center of nearly all massive galaxies and accrete nearby material. These are one of the most powerful sources classes in the Universe, and emit over the entire electromagnetic spectrum. It has been shown that the masses of the SMBHs correlate with that of the host galaxy bulge, velocity dispersion, and luminosity \citep{Magorrian_1998, Richstone1998, Gebhardt_2000, Merritt_2001, Ferrarese2005, Kormendy2013}. This trend indicates that SMBHs may determine star formation rates, due to molecular and ionized outflows \citep{Ferrarese2000, Gebhardt_2000, DiMatteo2005, Merloni_2010, Fiore2017, MartinNavarro2018}. If true, then the cosmic evolution of SMBHs and their host galaxies are inextricably linked. Therefore, being able to study the properties of SMBHs, including the gas and dust that surrounds them, becomes crucial. \\
\indent One of the best ways to study AGN through cosmic time is the cosmic X-ray background (CXB), i.e., the diffuse X-ray emission from 1  to 200--300\,keV \citep[e.g.,][]{Alexander_2003, Gilli2007, Treister_2009, Ueda_2014, Brandt2021CXB}. Models have shown that a significant fraction \citep[15-20\%;][]{Gilli2007, Ananna2019} of the peak of the CXB \citep[$\sim$30\,keV,][]{Ajello_2008} is generated by a population of AGN with large obscuring column densities, N$_{\rm H,los}$ $\geq$ 10$^{24}$ cm$^{-2}$, labeled as Compton-thick (CT-) AGN. Additionally, population synthesis models designed to accurately describe the origins of the CXB estimate that between 20\% \citep{Ueda_2014} and 50\% \citep{Ananna2019} of all AGN are CT. Nonetheless, the current fraction of observed CT-AGN is only between 5\% and 10\%, even at low redshifts \citep[i.e., z $<$ 0.01;][]{Burlon2011, Ricci_2015, TorresAlba2021}. \\
\indent CT-AGN are challenging to discover because the majority of their emission, from the optical through the soft X-rays, is obscured by the surrounding dust and gas \citep[i.e., the torus;][]{Urry1995}. However, the hard X-rays ($>$10\,keV) and the mid-infrared (MIR, 3--30$\mu$m) are able to pierce through the torus up to high column densities, making them the least biased bands against the detection of heavily obscured AGN \citep{Treister2004, Stern2005, Alexander2008}. The hard X-ray emission is created when UV light from the accretion disk interacts with hot electrons in the corona above the disk, thus Compton up-scattering into the hard X-ray band \citep{Haardt1993}. Additionally, the same UV radiation is absorbed by the dust, which in turn emits thermally in the IR \citep{Almeida2017, Honig2019}. Because of this, the emission in these two bands is expected to correlate significantly in AGN. Therefore, targeting the X-rays and the MIR is the ideal way to discover new CT-AGN. \\
\indent Observing and analyzing spectra of AGN in the X-rays and infrared to identify strong CT candidates is a time and resource intensive endeavor \citep[see, e.g.,][]{Marchesi2017, Andonie2022, Silver2022}. The Burst Alert Telescope (BAT) on board \textit{Swift} \citep{Gehrels2004} detected 1390 sources in its 150-month catalog\footnote{\url{https://science.clemson.edu/ctagn/bat-150-month-catalog/}}, almost 500 more than its predecessor, the 100-month catalog. With the addition of hundreds of sources in every catalog release, an efficient and accurate method to identify potential heavily obscured AGN is necessary. For this reason, our team has developed a new multiple linear regression machine learning algorithm to predict the line-of-sight column density of AGN. We have constructed a large sample of AGN with known N$_{\rm H}$ values and trained the algorithm using their MIR from the \textit{Wide-field Infrared Survey Explorer} \citep[WISE,][]{Wright2010}, soft X-ray data from the \textit{Swift}-X-ray Telescope (XRT), and hard X-ray data from \bat, to accurately predict the column density of new AGN. This work proceeds as follows: Section \ref{sec:sample} discusses the creation of our sample while Section \ref{sec:data_analysis} describes how the N$_{\rm H}$ values were determined. Section \ref{sec:machlearn} details the algorithm implemented and the input parameters included. Section \ref{sec:results} discusses the results of our algorithm and compares our predictive capabilities to other recent methods based on linear data modeling, rather than on multi-parameter machine learning algorithms like what is presented in this paper.

\section{Sample Selection} \label{sec:sample}
Our sample is taken from the 1390 sources detected in the BAT 150 Month catalog\footnote{The online version of the catalog can be found at: \url{https://science.clemson.edu/ctagn/bat-150-month-catalog/}} (Imam et al. in preparation). Of these 1390 sources, 568 are AGN with reliable N$_{\rm H}$ determinations (see Section \ref{sec:data_analysis} for details). \cite{Asmus2015} showed that the ratio of the MIR and X-ray flux can be a strong predictor of column density. Therefore, our machine learning algorithm (see Section \ref{sec:params}) includes data from XRT and WISE, so we cross-matched (using the BAT counterpart coordinates) with the 2SXPS \citep{Evans_2020} and AllWISE \citep{Cutri2014} with 5$\arcsec$ and 10$\arcsec$, respectively. For the 2SXPS and AllWISE, we found an average separation of $\sim$1.7$\arcsec$ and $\sim$1.8$\arcsec$, and a standard deviation of $\sim$1.1$\arcsec$ and $\sim$1.5$\arcsec$, respectively. This left us with a sample of 451 sources to train and test our machine learning algorithm (see Section \ref{sec:machlearn}).

\section{Data Analysis} \label{sec:data_analysis}
The majority of the sources (361) in our sample of 451 are in the BAT 70-month catalog \citep{Ricci2017}, which provides N$_{\rm H}$ values based on spectral analysis of soft X-ray (\textit{ASCA}, \textit{Chandra}, \textit{Suzaku}, \textit{Swift}-XRT, and \textit{XMM-Newton}) and BAT spectra. For the remaining 90 sources, we modeled their soft X-ray jointly with their \textit{Swift}-BAT spectra. \xmm data was available for 18 sources, while \cha data was available for an additional 24. For the remaining 48 sources, the soft X-ray data were provided by \xrt. As the greater part of the sources in the sample were unobscured (Log(N$_{\rm H}) <$ 22) or mildly obscured (22 $<$ Log(N$_{\rm H}) <$ 23), they were sufficiently modeled with an absorbed powerlaw, as shown below:

\begin{multline}
    Model1 = constant_1 * phabs * (zphabs * zpowerlw),
\end{multline}

However, Compton-thin (23 $<$ Log(N$_{\rm H}) <$ 24) sources required a more complex model to account for the Fe K$\alpha$ emission and the fraction of intrinsic emission that leaks through the torus rather than being absorbed by the obscuring material. These sources were modeled as such:

\begin{multline}
    Model2 = constant_1 * phabs * (zphabs * zpowerlw \\ 
    + zgauss + constant_2 * zpowerlw),
\end{multline}

where $constant_1$ accounts for cross-normalization differences between the soft X-ray instrument and \bat, $phabs$ models the galactic absorption, $zphabs * zpowerlw$ is the absorbed power-law modeling the intrinsic emission, $zgauss$ models the Fe K$\alpha$ emission line, and $constant_2 * zpowerlw$ represents the scattered emission that leaks through the torus. \\
\indent When sources approach or surpass the Compton-thick limit, they require even more sophisticated modeling. These sources were modeled with physically motivated models such as \texttt{MYTorus} \citep{Murphy2009} and \texttt{borus02} \citep{Balo_borus2018}, and have been described in detail in \cite{Zhao2019a, Zhao2019b, TorresAlba2021, Silver2022}. These models are used for heavily obscured AGN because they account for the photons that interact with the dust and gas surrounding the SMBH and are reflected into the observer line of sight.

\section{Machine Learning} \label{sec:machlearn}
\subsection{Multiple Linear Regression} 
Linear regression is one of the most commonly used machine learning techniques \cite[see, e.g.,][]{Chen2021, Mizukoshi2022}. Simply, linear regression models the linear relationship between an explanatory variable (input parameter) and the response variable (output parameter). Since few quantities can be accurately modeled using only one explanatory variable, using numerous can improve the predictive capability of an algorithm. This is referred to as multiple linear regression, or just multiple regression for short, and is modeled as shown below: 

\begin{equation}
    y = \beta_0 + \beta_1 x_1 + \beta_2 x_2 + ... \beta_i x_i,
\end{equation}

where y is the response variable, $x_i$ are the explanatory variables, $\beta_0$ is the y-intercept (if necessary), and $\beta_i$ are the slope coefficients corresponding to each explanatory variable. Using a large sample of sources with data for every explanatory variable and a known value for the response variable, the algorithm trains itself to determine which combination of $\beta_i$ values is optimal for reproducing the response variable. Out of our sample of 451 sources, 80\% were randomly selected to be used in our training sample, thus leaving 20\% (91 sources) left for our test sample. We determined this was the optimal ratio as it used enough sources to accurately train the algorithm while simultaneously leaving a statistically significant sample to verify this accuracy \footnote{We note that neural networks are another commonly used machine learning algorithm \citep[see, e.g.,][]{Finke2021, Chainakun2022, Zubovas2022}. We applied this technique to our data set and after finding the optimal configuration, yielded very similar results to those generated by our linear regression model. For this reason, we have optioned to present the results from the comparatively simpler linear regression model in this paper.}. \\

\subsection{Parameters Used} \label{sec:params}
The algorithm will only be as accurate as however strong the relationship is between the chosen input parameters and the desired output parameter. We have selected WISE colors, BAT count rates, soft X-ray hardness ratios (HRs), and an MIR$-$soft X-ray flux ratio as all have been previously shown to correlate with N$_{\rm H}$. These parameters are described below.

\subsubsection{MIR Colors}
Roughly half of the intrinsic emission from the AGN is absorbed by the dusty torus \citep[see, e.g.,][]{Almeida2017, Honig2019}. As a consequence, the dust present in the torus is heated to temperatures of several hundred Kelvin, and thus radiates thermally. This emission peaks in the MIR ($\sim$3--30$\mu$m) and is much less prone to absorption than the optical and UV, making it a crucial tool to study obscured AGN. With the launch of WISE, we have an all-sky instrument with superb resolution ($\sim$6'') capable of studying these obscured sources. WISE observes the entire sky in four bands, 3.4, 4.6, 12, and 22\,$\mu$m (W1, W2, W3, and W4, respectively), and to date has detected nearly 750 million sources and reached flux limits of 7.1$\times$10$^{-14}$ erg cm$^{-2}$ s$^{-1}$. The differences between these bands have been proven to be a good predictor for different levels of obscuration. \cite{KilerciEser2020} used a sample of AGN from the BAT 105-month catalog \citep{Oh2018} to create new CT-AGN selection criteria based on MIR colors. They find that the median values for different colors have an increasing trend with N$_{\rm H}$ (see Figure 9 from their paper). Therefore, our algorithm includes six WISE colors: W1-W2, W1-W3, W1-W4, W2-W3, W2-W4, and W3-W4.

\subsubsection{MIR--X-ray Flux Ratio}
\begin{figure*} 
    \centering
    \includegraphics[scale=0.8]{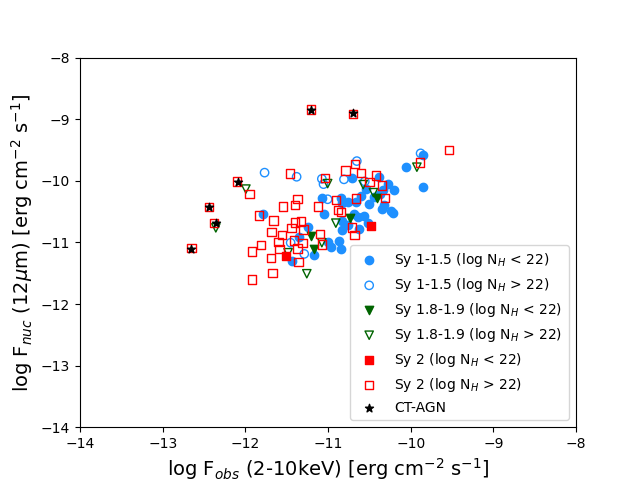}
    \caption{The relationship between the observed 12$\mu$m flux and the observed 2--10\,keV flux, adapted from Figure 1 in \cite{Asmus2015}. The closed shapes represent unobscured AGN (LogN$_{\rm H}<$22) while open shapes represent obscured AGN (LogN$_{\rm H}>$22). The blue circles are Seyfert (Sy) 1-1.5 galaxies; green triangles are Sy 1.8-1.9 galaxies; and red squares are Sy 2 galaxies. The black stars represent confirmed CT-AGN. The obscured sources fall to the left of the trend, signifying that the ratio between these two quantities can be a tracer of obscuration.}
    \label{fig:asmus15}
\end{figure*}

As the X-ray and MIR emission are both reprocessed from the same material, it is expected that a correlation exists between them. \cite{Asmus2015} shows the trend between the observed 12\,$\mu$m flux and the 2--10\,keV flux (see Figure \ref{fig:asmus15}). Moreover, a shift is evident in the trend based on obscuration. As source obscuration increases, the observed 2--10\,keV flux decreases, thus causing the source to fall to the left of the predicted trend. This is evidenced in the figure by Seyfert 2 galaxies (red squares) falling to the left of traditionally unobscured Seyfert 1 galaxies (blue circles). Moreover, the confirmed CT-AGN (black stars) fall well to the left of even Seyfert 2 galaxies, suggesting an extremely suppressed observed X-ray flux. As a result of this trend, \cite{Asmus2015} used the log ratio of the 12\,$\mu$m flux density and the 2--10\,keV flux to predict the column density of an AGN. We have included this parameter in our algorithm, using the 12\,$\mu$m flux density measurement from WISE and the 2--10\,kev flux from \textit{Swift}-XRT.

\subsubsection{Soft X-ray Hardness Ratios}
\begin{figure*}
    \centering
    \includegraphics[scale=0.8]{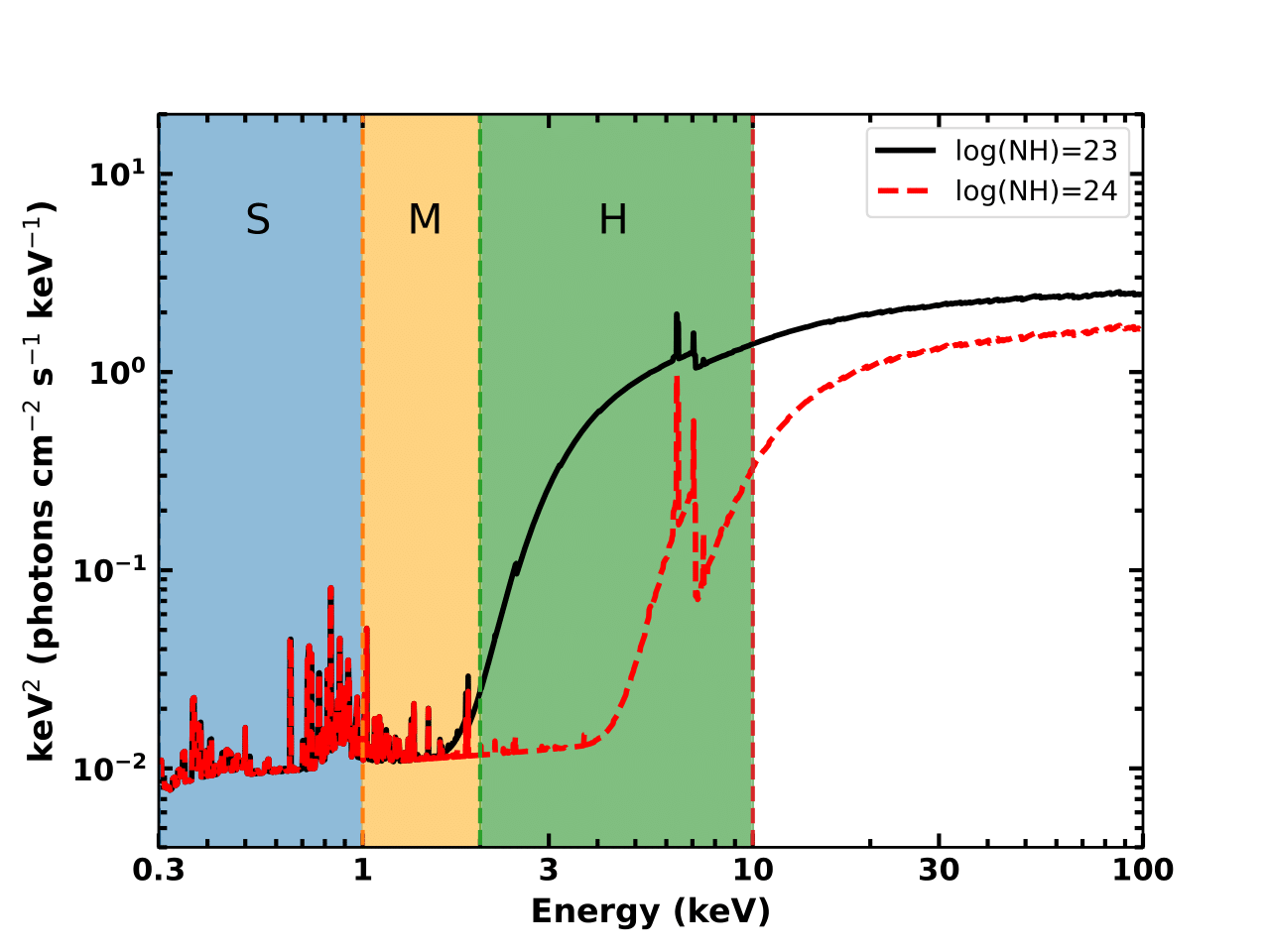}
    \caption{Simulated X-ray spectra of an AGN with line-of-sight column density Log(N$_{\rm H}$)=23 (black solid line) and Log(N$_{\rm H}$)=24 (red dotted line). The vertical regions denoted by S (blue), M (orange), and H (green) represent the different bands used in our two hardness ratios where they correspond to the 0.3--1, 1--2, and 2--10\,keV bands, respectively. The two spectra show extreme differences in the soft X-rays, particularly in the 2--10\,keV band. Thus, hardness ratios targeting this band are helpful in determining the column density of AGN.}
    \label{fig:nh23_24}
\end{figure*}

Soft X-rays (0.3--10\,keV) are very susceptible to changes in column density, as evidenced in Figure \ref{fig:nh23_24}. It can be seen that the 0.3--10\,keV emission is far more suppressed in a source with Log(N$_{\rm H}$)=24 compared to a source with Log(N$_{\rm H}$)=23. Therefore, the ratio between the counts in different energy bands, or hardness ratios, covering this energy band are highly dependent on column density. For this reason, we have included two hardness ratios from the latest \textit{Swift}-XRT point source catalog, the 2SXPS \citep{Evans_2020}; (M-S)/(M+S) and (H-M)/(H+M) where S, M, and H correspond to the 0.3--1, 1--2, and 2--10\,keV bands.

\subsubsection{Hard X-ray Count Rates}
\begin{figure*}
    \centering
    \includegraphics[scale=0.8]{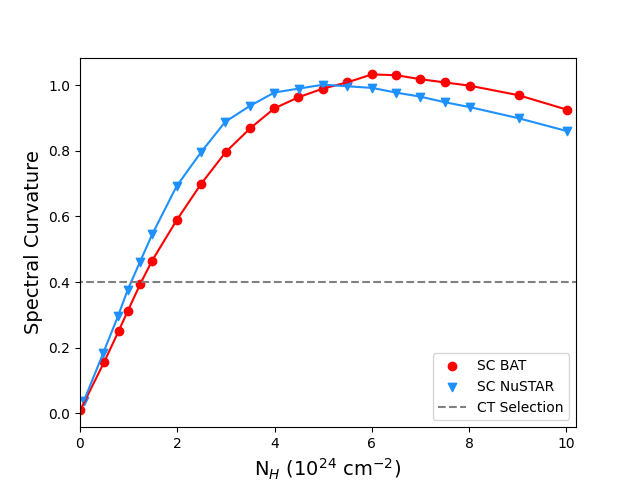}
    \caption{The spectral curvature value based on different column densities, displaying some of the configurations adapted from Figure 3 in \cite{Koss2016}. Each curve represents how the SC value changes with N$_{\rm H}$ based on different input parameters used in their \texttt{MYTorus} model simulations. The red circles show the curve when the SC equation is calibrated to \textit{Swift}-BAT data, while the blue triangles show the different SC equation when calibrated to \textit{NuSTAR} data. The dashed grey line indicates the cutoff for CT-AGN determined by \cite{Koss2016}. Both lines illustrate how the increased curvature of hard X-rays of AGN is related to an increase in column density.}
    \label{fig:koss16_sc}
\end{figure*}

While significantly less affected than soft X-rays, hard X-rays do display an increased curvature with higher column densities. \cite{Koss2016} analyzed sources with \textit{Swift}-BAT data and found a correlation between the spectral curvature and column density. Using simulated data of CT-AGN, they generated the following equation:

\begin{equation} \label{eq:bat_sc}
    \text{SC}_{\text{BAT}} = \frac{-3.42 \times A - 0.82 \times B + 1.65 \times C + 3.58 \times D}{\text{Total Rate}},
\end{equation}

where A, B, C, and D refer to the 14--20\,keV, 20--24\,keV, 24-35\,keV, and 35--50\,keV bands, while the total rate is the 14--50\,keV band. As plotted in Figure \ref{fig:koss16_sc}, an increase in this value (calculated via Equation \ref{eq:bat_sc}) was linked to an increase in line-of-sight column density. Two different models that calculate N$_{\rm H}$ are plotted and all agree that the spectral curvature value increases with N$_{\rm H}$. However, we note that this method is only valid up to N$_{\rm H}$ = 4$\times$10$^{24}$ cm$^{-2}$. \\
\indent We used this principle to improve our algorithm. Whereas \cite{Koss2016} only included data up to 50\,keV, we found our algorithm performed better when including data up to 150\,keV. Because of this, we included BAT count rates for nine different energy bands in our algorithm: 14--20\,keV, 20--24\,keV, 24--34\,keV, 34--45\,keV, 45--60\,keV, 60--85\,keV, 85--110\,keV, 110--150\,keV, and 14--150\,keV. Including each band as a parameter accounts for both the curvature in the spectrum while also serving as a proxy for the BAT flux. For this reason, we elected to include every band instead of just the spectral curvature value.

\section{Results} \label{sec:results}
\begin{figure*}
    \centering
    \includegraphics[scale=0.8]{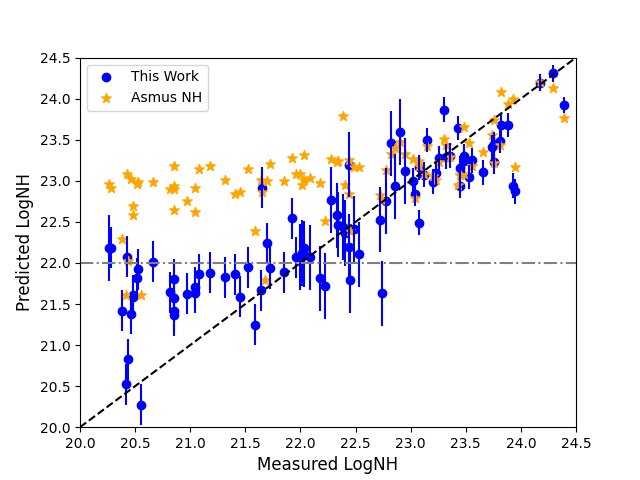}
    \caption{The x-axis shows the ``true'' line-of-sight Log(N$_{\rm H}$) values, as determined by spectral fitting. The y-axis shows the Log(N$_{\rm H}$) values predicted by our machine learning algorithm (blue circles) and those predicted by the \cite{Asmus2015} equation (orange stars). 
    Our algorithm shows superior predictive capabilities, particularly for lower levels of obscuration (Log(N$_{\rm H}) <$ 23), where our algorithm does not incorrectly classify unobscured sources as heavily obscured as displayed by the grey dash-dotted line. The black dotted line represents the one-to-one ratio between the ``true'' and predicted N$_{\rm H}$ values. The errors from our algorithm were calculated statistically. No errors are included on the orange points for readability purposes.}
    \label{fig:ml_vs_asm}
\end{figure*}

\begin{figure*}
    \centering
    \includegraphics[scale=0.8]{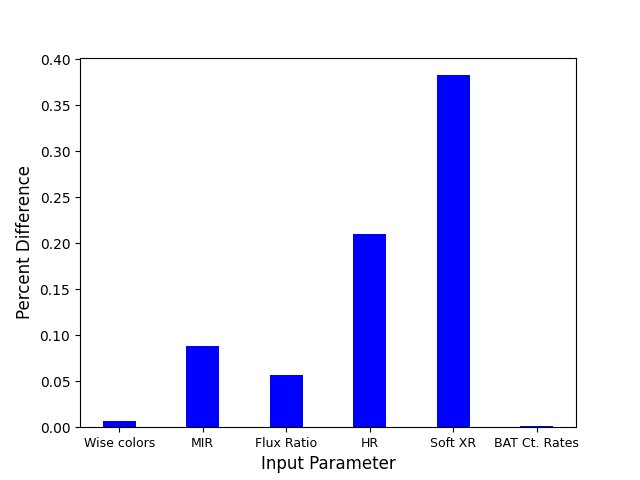}
    \caption{The percent difference between the Spearman correlation coefficient including all parameters and the coefficient when the listed parameter is excluded. The larger the difference, the worse the fit without that parameter (i.e. the higher the importance of that parameter). Therefore, the soft X-ray-related parameters are the highest contributors to the predictive capability of the algorithm. ``MIR'' refers to the WISE colors and the MIR-X-ray flux ratio. ``HR'' represents the two X-ray hardness ratios. ``Soft XR'' refers to the two X-ray hardness ratios and the MIR-X-ray flux ratio.}
    \label{fig:feat_import}
\end{figure*}

\indent Figure \ref{fig:ml_vs_asm} shows the X-ray-confirmed N$_{\rm H}$ values for the 91 sources plotted against the predictions by our algorithm (blue circles). We used the Spearman rank correlation coefficient to measure the strength of the correlation between the two sets of N$_{\rm H}$ values. Our algorithm yielded a Spearman coefficient of 0.86, signifying that it performs very well in recreating the true N$_{\rm H}$ values of these sources. Moreover, of the 31 heavily obscured (Log(N$_{\rm H}$) $\geq$ 23) sources in our test sample, our algorithm correctly predicted 25 of them (80\%) with Log(N$_{\rm H}$) $>$  23 and 30 (97\%) with Log(N$_{\rm H}$) > 22.80. We note that we are currently unable to distinguish this obscuration as being caused by the nuclei or the host galaxy, particularly for edge-on galaxies. \\
\indent In order to determine which input parameters were most impactful in training our algorithm, we used the percent difference of the Spearman correlation coefficient when all parameters were used (0.86) and the coefficient when only the parameter listed was excluded. The larger this difference, the worse our algorithm performed without including said input parameter. Since removing one WISE color or BAT count rate had little effect, we grouped the parameters as such: all six WISE colors; WISE colors + the MIR-X-ray flux ratio (MIR); the MIR-X-ray flux ratio; the two XRT hardness ratios; the two hardness ratios + the MIR-X-ray flux ratio (Soft X-ray); and the BAT count rates. Since the MIR-X-ray flux ratio includes both information from the infrared and the X-rays, we included separate categories without it to determine which wavelength influenced our algorithm the most. Figure \ref{fig:feat_import} shows that the three parameters using soft X-ray data (the two XRT hardness ratios and the MIR-X-ray flux ratio) were the largest contributors towards our algorithm producing accurate results.

\subsection{Comparison with Previous Methods}

\subsubsection{\cite{Asmus2015}}
Utilizing MIR data alongside soft X-ray data of a sample of 152 AGN, \cite{Asmus2015} determined a relation to predict line-of-sight column densities. The relation is as follows: 

\begin{multline}
\text{log} \left(\frac{N_\text{H}}{\text{cm}^{-2}}\right) = (14.37 \pm 0.11) + (0.67 \pm 0.11) \\ 
\times \text{log}\left(\frac{F^{\text{nuc}} (12\mu \text{m})}{F^{\text{obs}} (2 - 10\,\text{keV})} \frac{\text{erg s$^{-1}$ cm$^{-2}$}}{\text{mJy}} \right).
\end{multline}

Using the WISE 12\,$\mu$m and XRT 2--10\,keV fluxes, we plotted the N$_{\rm H}$ values predicted by the Asmus relation for our test sample of 91 sources in Figure \ref{fig:ml_vs_asm} . While these results show a good trend for heavily obscured sources, below Log(N$_{\rm H})$ = 23, our machine learning algorithm performs far better. This is quantified by the lower Spearman correlation coefficient of 0.65 for the Asmus predictions and the real N$_{\rm H}$ values. The lack of predictive capability below 10$^{23}$ cm$^{-2}$ affects the whole range of possible N$_{\rm H}$ values. This is because, a priori, one does not know the `true' N$_{\rm H}$ of the source, and if choosing a source with Log(N$_{\rm H}) <$ 23, the relation will confidently place it as being heavily obscured. Therefore, sources with Log(N$_{\rm H}) <$ 23 can actually have any value of `true' Log(N$_{\rm H}$) between 20 and 23.

\subsubsection{\cite{Pfeifle2022}}
\cite{Pfeifle2022} improved upon the work of \cite{Asmus2015} by creating a new relationship based on the ratio of the 2--10\,keV and 12\,$\mu$m luminosities. Using 456 AGN detected in the 70-month BAT catalog \citep{Ricci2017} that also possess infrared data, their team created the relation listed below:

\begin{multline}
\text{log} \left(\frac{N_\text{H}}{\text{cm}^{-2}}\right) = 20 + (1.61^{+0.33}_{-0.31})  \\ 
\times \text{log}\left(\Bigg | \frac{\text{log}(\frac{L_{X,Obs.}}{L_{12\mu \text{m}}}) + (0.34^{+0.06}_{-0.06})} {(-0.003^{+0.002}_{-0.005})}\Bigg | \right).
\end{multline}

\begin{figure*}
    \centering
    \includegraphics[scale=0.8]{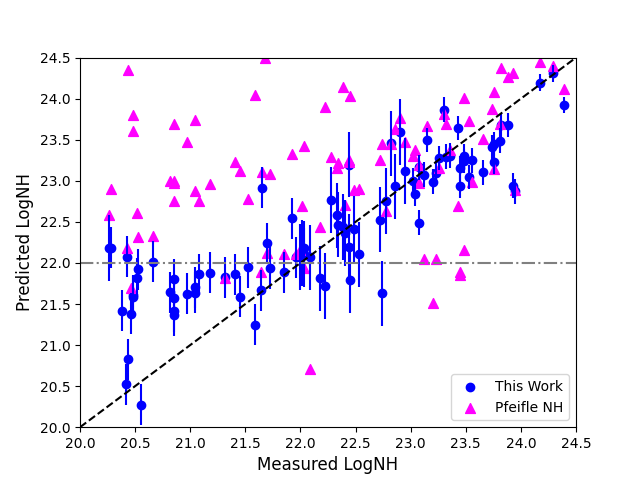}
    \caption{As in Figure \ref{fig:ml_vs_asm}, the x-axis shows the ``true'' line-of-sight Log(N$_{\rm H}$) values determined by spectral fitting while the y-axis shows the Log(N$_{\rm H}$) values predicted by our machine learning algorithm (blue circles) and by the \cite{Pfeifle2022} relation (magenta triangles). The errors from our algorithm were calculated statistically. No errors are included on the magenta points for readability purposes.}
    \label{fig:pfeifle}
\end{figure*}

With this relation, we have predicted the N$_{\rm H}$ values for our sample of 91 test sources as seen as the magenta triangles in Figure \ref{fig:pfeifle}. \cite{Pfeifle2022} claims that their method is most accurate when applied to sources with Log(N$_{\rm H}) >$ 22.5, which is confirmed here. While their method is accurate for heavily obscured sources, it is far less predictive than our algorithm for AGN with Log(N$_{\rm H}) <$ 22.5. Just as with the Asmus relation, this represents a significant drawback when selecting sources as we do not know whether or not the `true' Log(N$_{\rm H}) >$ 22.5. Overall, it has a Spearman correlation coefficient of 0.27.

\subsubsection{\cite{Koss2016}} \label{sec:koss_compare}
\begin{figure*}
    \centering
    \includegraphics[scale=0.8]{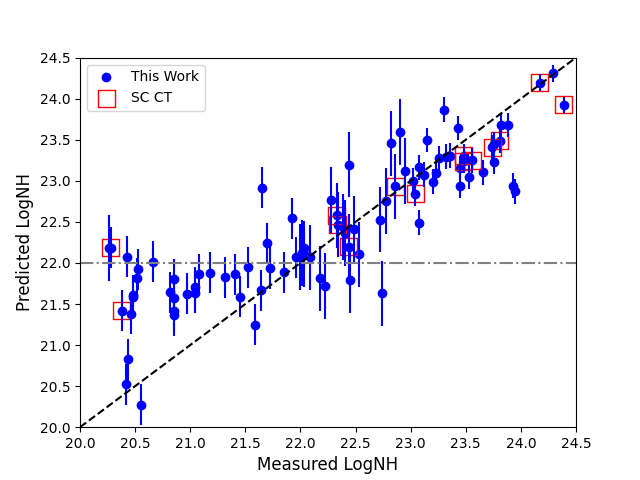}
    \caption{As in Figure \ref{fig:ml_vs_asm}, the x-axis shows the ``true'' line-of-sight Log(N$_{\rm H}$) values determined by spectral fitting while the y-axis shows the Log(N$_{\rm H}$) values predicted by our machine learning algorithm (blue circles). The red squares represent the 14 sources that were predicted to be CT based on the spectral curvature method introduced in \cite{Koss2016}. 6 of these sources (43\%) have true N$_{\rm H}$ values $<$ 10$^{23}$ cm$^{-2}$. Our algorithm makes no such misclassifications. The errors from our algorithm were calculated statistically.}
    \label{fig:ml_koss}
\end{figure*}

\cite{Koss2016} developed a method to identify new CT-AGN using weighted averages of different \bat bands. It was determined that an SC$_{BAT} >$ 0.40 would identify a CT-AGN candidate. We applied this formula to our 91 test sources and found 14 that would be considered CT. These sources are plotted as red squares in Figure \ref{fig:ml_koss}, overlapped on our machine learning predictions. We note that this method does show promise, as 8 of the 14 sources are heavily obscured, with Log(N$_{\rm H}) >$ 23. However, 6 sources (43\%) predicted as CT have true Log(N$_{\rm H}) <$ 23, including two that are unobscured (Log(N$_{\rm H}) <$ 22). Additionally, of the 14 predicted as CT, only 2 (14\%) truly are. Our machine learning algorithm does not misclassify any unobscured sources as CT and performs more accurately throughout all column density ranges. Moreover, both sources predicted as CT by our algorithm are truly Compton-thick. 

Table \ref{tab:nh_class} displays how many of the 91 test sources are divided in each of these four categories: Compton-thick (Log(N$_{\rm H}) >$ 24), Compton-thin (23 $<$ Log(N$_{\rm H}) <$ 24), obscured (22 $<$ Log(N$_{\rm H}) <$ 23), and unobscured (Log(N$_{\rm H}) <$ 22). As can be seen, our machine learning algorithm performs the best overall, correctly classifying $\sim$75\% of the sources. This is particularly true for sources with Log(N$_{\rm H}) <$ 23, in which our algorithm has a 73\% accuracy, while the other two applicable methods are both only 18\% accurate.

\section{Conclusions}
In this work, we present a new machine learning algorithm that predicts the line-of-sight column density of AGN, thus enabling us to discover new CT-AGN candidates. Using MIR data from WISE, soft X-ray data from \textit{Swift}-XRT and hard X-ray data from \textit{Swift}-BAT, our machine learning algorithm has proven its ability to accurately reproduce the N$_{\rm H}$ values of our 91-source test sample, correctly classifying 75\% of sources based on their obscuration. Moreover, our algorithm has shown a superior ability to predict the column density of AGN with Log(N$_{\rm H}$) $<$ 22.5 when compared with previously published methods. In the future, this algorithm will be used to: 1) identify promising CT-AGN candidates and 2) efficiently determine N$_{\rm H}$ values of large samples of sources (like the \textit{Chandra} and \xmm source catalogs) in an effort to determine the obscuration distribution of the entire AGN population across cosmic time.

\begingroup
\renewcommand*{\arraystretch}{2}
\begin{table*}
    \centering
    \caption{We have split the 91 test sources into four classifications based on their X-ray-measured column densities: Compton-thick (Log(N$_{\rm H}) >$ 24), Compton-thin (23 $<$ Log(N$_{\rm H}) <$ 24), obscured (22 $<$ Log(N$_{\rm H}) <$ 23), and unobscured (Log(N$_{\rm H}) <$ 22). The number of sources correctly classified for each of the four methods mentioned in this paper are shown below.}
    \label{tab:nh_class}
    \begin{tabular}{lcccccc}
    \hline \hline
    \textbf{Classification} & \textbf{Real Number} & \textbf{This Work} & \textbf{\cite{Asmus2015}} & \textbf{\cite{Pfeifle2022}} & \textbf{\cite{Koss2016}} \\
    \hline
    Compton-thick & 3 & 2 & 2 & 3 & 2 \\
    Compton-thin & 28 & 22 & 25 & 13 & ... \\
    Obscured & 24 & 16 & 8 & 7 & ... \\
    Unobscured & 36 & 28 & 3 & 4 & ... \\
    \hline
    \textbf{Total} & 91 & 68 (75\%) & 38 (42\%) & 27 (30\%) & ... \\
    \hline \hline
    \end{tabular}
\end{table*}
\endgroup

\bibliographystyle{aa}
\bibliography{bibliography}

\end{document}